\documentclass[12pt,preprint2]{aastex}

\shorttitle{Evidence for  the naphthalene cation in the ISM }
\shortauthors{S. Iglesias-Groth et a.}

\begin{document}

\title{Evidence for the  naphthalene cation  in a region of the interstellar medium with anomalous microwave emission} 

\author{S. Iglesias-Groth, A. Manchado\altaffilmark{1} and D. A. Garc\'\i
a-Hern\'andez} 
\affil{Instituto de Astrof\'\i sica de Canarias, C/ Via L\'actea s/n, 38200 La
Laguna, Spain}
\email{sigroth@iac.es}
\author{J. I. Gonz\'alez Hern\'andez\altaffilmark{2}}
\affil{Observatoire de Paris, GEPI, 92195 Meudon Cedex, France} 

\and

\author{ D.L.  Lambert}
\affil{The W.J. McDonald Observatory, University of Texas, Austin, TX 78712-1083, USA}          

\altaffiltext{1}{Consejo Superior de Investigaciones Cient\'\i ficas, Spain}
\altaffiltext{2}{CIFIST Marie Curie Excellence Team}

\begin{abstract}
  We report high resolution spectroscopy of the moderately reddened (A$_V$=3) early type star Cernis 52   located in a region of the Perseus molecular cloud complex with anomalous microwave emission. In addition to the presence of the most common diffuse interstellar bands (DIBs) we detect two new interstellar or circumstellar bands coincident to within 0.01\%  in wavelength with the two strongest  bands of the naphthalene cation (C$_{10}$H$_{8}^+$) as measured in gas-phase laboratory spectroscopy at low temperatures and find  marginal evidence for the third strongest band.
  Assuming these features are caused by  the naphthalene cation, from the measured intensity and available oscillator strengths we find that 0.008 \% of the carbon in the cloud  could be in the form of this molecule.  We expect hydrogen additions to cause hydronaphthalene cations  to be abundant in the cloud and to contribute via electric dipole radiation to the anomalous microwave emission. The  identification of new interstellar features consistent with transitions  of  the simplest polycyclic aromatic hydrocarbon adds support to the hypothesis that this type of molecules are the carriers of both diffuse interstellar bands and  anomalous microwave emission.

\end{abstract} 
  
\section{Introduction}

 Polycyclic aromatic hydrocarbons (PAHs)   have been postulated as the
 natural  carriers for: a) the  mid-infrared emission features widely
 observed in the  interstellar medium (Puget \& L\'eger 1989;
 Allamandola, Tielens \& Barker 
 1985, 1989); b) the diffuse interstellar bands (DIBs), hundreds of 
 absorption bands in the optical and near-infrared  spectra of stars
 reddened by interstellar  material (L\'eger \& d\'~Hendecourt 1985,
 Van der Zwet \& Allamandola 1985) which despite decades of
 observations remain as  one of the most enigmatic problems faced by
 stellar spectroscopists (Herbig 1995, Krelowski et al. 2001, Mulas et
 al.  2006, Sarre 2006, Hobbs et al. 
 2008); and c)   the so-called anomalous microwave emission, a diffuse
 emission  detected at high  Galactic latitude by several cosmic
 microwave experiments  (Kogut et al. 1996, Leitch et al. 1997, de
 Oliveira et al. 1999, 2002;  Hildebrandt et al. 2007) possibly
 produced by electric dipole radiation of rapidly spinning
 carbon-based molecules (Draine \& Lazarian 1998; Iglesias-Groth 2005,
 2006). 

We decided to explore a possible connection between the carriers of the last two phenomena by performing observations of DIBs towards the Perseus   molecular cloud
 complex where recent observations by Watson et al. (2005) demonstrated the existence of regions of anomalous microwave emission. The microwave interferometer VSA
 (Very Small Array, Taylor et al.2003) has recently shown a correlation between the angular structure of the anomalous  emission in Perseus and the thermal dust
 emission measured by the IRAS satellite at angular scales of several arcminutes (Watson et al. 2006).We  searched for DIBs in the line of sight of the maximum
 anomalous microwave emission detected by the VSA by performing high resolution optical spectroscopy of Cernis 52  (BD+31$^o$ 640), a reddened star (excess colour
 E(B-V)=0.9, Cernis 1993) located in this line of sight at distance of 240 pc, where most of the dust extinction is  known to concentrate in the Perseus OB2 dark
 cloud complex. This is the brightest early type star in the region of interest. We identify numerous DIBs towards Cernis 52 and report here on the detection of
 two new  bands of likely interstellar origin with wavelengths and strengths consistent with those of the  naphthalene cation (C$_{10}$H$_{8}^+$). Naphthalene  is
 the simplest PAH, its cation has been  searched in the interstellar medium (Snow et al. 1992, Kre{\l}owski et al. 2001) without a conclusive identification,
 partly due to the lack of accurate  wavelengths for resonance transitions of the cation. Recent gas phase  laboratory spectroscopy at low temperatures  provide
 precise wavelengths, widths and oscillator strengths  (Pino et al. 1999, Romanini et al. 1999, Biennier et al. 2003) and may allow a proper identification of this cation in space.

\begin{figure}
\includegraphics[angle=0,width=8.5cm]{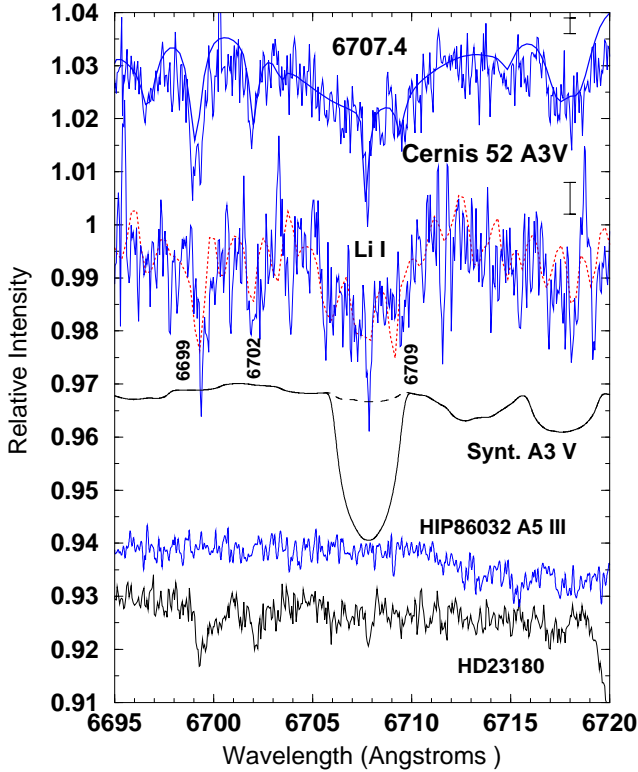}
\caption{Spectra of star Cernis 52  showing three independent detections of the new 6707.4 \AA~band. The two spectra at the top (blue thin lines) were obtained  with the 2.7 m
telescope at Mc Donald Observatory in different epochs (see text). 1$\sigma$  error bars are indicated. At the top, we overplot (thick blue line) a  fit to the data resulting
from a combination of  a Lorentzian profile of width 12 \AA~  for this band with  suitable profiles for known narrower DIBs and a synthetic photospheric spectrum assuming cosmic
Li abundance. The  red line in the second spectrum shows the lower resolution  data obtained at the 4.2m  WHT. The narrow feature at 6707.7 \AA~ is   associated with
interstellar lithium.  The known  DIBs at 6699.32, 6702.02 and 6709.43 \AA~ are marked with wavelength  and  can be also identified in our spectrum  of the hotter reference star
 HD 23180  (displayed at the bottom).  A synthetic spectrum (Synt A3 V) computed using a suitable model  atmosphere for Cernis 52 (see text for description)  is plotted  for two
 values of Li abundance   $\log [N({\rm Li})/N({\rm H})]+12 =3.3$ and 4.2 , respectively.  The synthetic spectra were convolved assuming a rotational velocity of 80 km/s, adequated for the star.  The weak broad feature at 6717.7  \AA~ is due to photospheric Ca. Star HIP 86032 (from Allende-Prieto et al.  2004) is   plotted for comparison  }

\end{figure}

\section{Observations}

  The data were obtained with the  SARG spectrograph (Gratton et al. 2001) on the 3.5 m Telescopio   Nazionale Galileo (TNG), the ISIS spectrograph at the 4.2 m William Herschel  Telescope (WHT), both  at the  Roque   de los Muchachos Observatory (La Palma,  Spain) and  the   2dcoud\'{e} cross-dispersed echelle (Tull et al.1995) on the   2.7 m Harlan J. Smith Telescope at McDonald Observatory (Texas, USA). 
       We  observed star Cernis 52  in November 2006 with SARG at resolving power  R=40000 in the range  5000-9000 \AA~and obtained  a final combined spectrum  with 
       S/N$\sim$100 in the continuum. In addition, star HD 23180 (Cernis 67, also known as o Per) was observed with the same configuration in a nearby line of sight
       selected with no significant anomalous microwave emission. Fast rotating early type stars were also observed  for correction of telluric lines. For control of
       any possible systematic effects  further spectroscopic data on Cernis 52 and HD 23180 were obtained on 28 December 2007  at WHT using ISIS with  grating 1200R,
       which provided a dispersion of 0.24 \AA/pixel and S/N of 220 in the continuum. Fifteen additional spectra of Cernis 52 where obtained on 23, 25, 26 August 2007, and   1, 2 March 2008 with the Mc Donald 2dcoud\'{e} cross-dispersed echelle spectrograph. The spectral range  3740-10160 \AA~ was recorded with  resolving power R=60000 and  R=30000, respectively.  These two  datasets  were  independently reduced and properly combined to produce final spectra of  S/N  $\sim$150 and 300,  respectively.  Early-type rapidly  rotating hot stars were also observed for correction of any possible telluric absorption in both campaigns. All the spectra were reduced using IRAF, wavelength calibrated with accuracy better than 20 m\AA~per pixel and co-added  after   correction for telluric lines. The zero-point of the adopted wavelength scale was set by assigning the  laboratory wavelengths to the Na D interstellar absorption lines recorded in our spectra.

\begin{figure}
\includegraphics[angle=0,width=8.5cm]{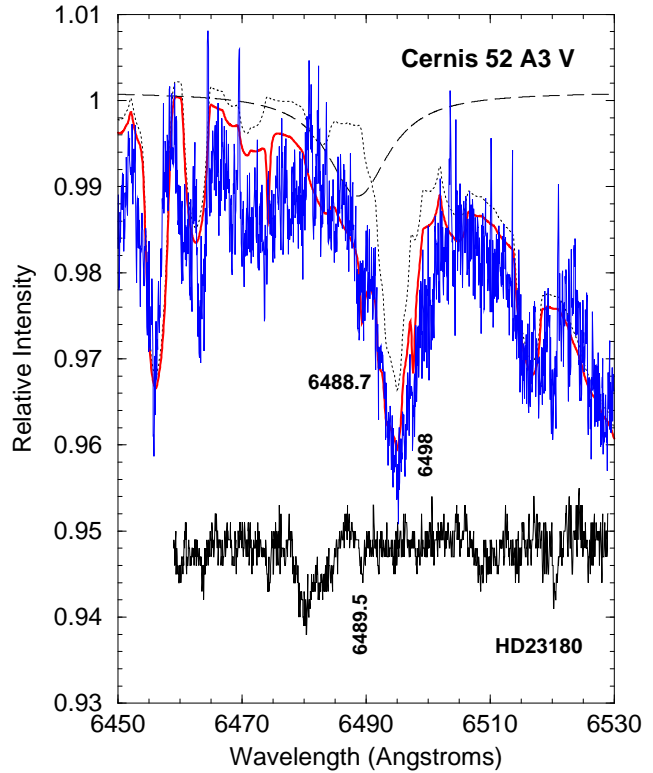}
\caption{Spectrum of the star Cernis 52 (thin blue  line) in the region 
of the second most intense laboratory band of the naphthalene cation.  
 For comparison we plot a synthetic photospheric
spectrum (dotted black line) computed using the same stellar  model as in the previous 
figure. The  strong absorption at 6495 \AA~ is caused by a 
combination of  photospheric and interstellar absorptions. Superimposed on 
the blue wing of this feature we find the   absorption at 6488.7 \AA~,  
which  we associate with the naphthalene cation. We plot (red line)  a combination of the synthetic spectrum with  a Lorentzian profile of width 12 \AA~ centered at this wavelength and bands at the positions of known DIBs. At the bottom we display  the spectrum of HD 23180 for reference (Galazutdinov et al. 2000). The DIBs at 6489.47  and 6498.08 \AA~ are marked.}
\end{figure}

\section{Discussion}

The star Cernis 52 appears embedded in or behind  a cloud responsible for significant visible extinction (A$_V$=3) and  absorption  bands caused by the intervening interstellar material were  expected in the spectrum.
 We  detected the strongest known DIBs (including  $\lambda\lambda$ 5780, 5797, 5850, 6196, 6270, 6284, 6376, 6379 and 6614 \AA) in the various sets of spectra. The  DIBs at 6270, 6376 and 6379 \AA~appear in Cernis 52 with normal strength for the amount of reddening. The  DIB at 6284 \AA~is however  remarkably weak. A strong UV field is needed to excite the carrier  of this  band (Kre{\l}owski $\&$ Walker 1987). This  strongly argues in favor of a UV shielded environment, a $\zeta$-type cloud (Kre{\l}owski and Sneden 1995),  as the  source for anomalous microwave emission.  The strength of the 5780 DIB is also consistent with such a type of cloud. The DIBs wavelengths show velocity shifts consistent with those of the Na D doublet. A complete study of the DIBs in the line of sight will be presented elsewhere.

Remarkably,  we  detected first in the TNG spectra and subsequently in each of the other spectra  a broad feature near 6707 \AA, which does not appear in compilations of DIBs (Jenniskens \& Dessert 1994;   Galazutdinov et al.  2000; Hobbs et al. 2008). In Fig. 1 we plot  the McDonald and WHT spectra  of Cernis 52    showing  clear  detections of this feature which is not present in either the nearby DIB reference star HD 23180, or the A5III star HIP 86032 (Allende-Prieto et al. 2004).  The closest known DIBs are located at 6699.32, 6702.02 and 6729.28 \AA~ (Hobbs et al 2008). We detect these three bands  in Cernis 52 with equivalent widths of 19, 9 and 10 m\AA, respectively (with an error of 20\%). They are also found in the comparison star HD 23180 (Galazutdinov et al 2000). A very narrow absorption feature at 6707.7 \AA~is also observed in the spectrum of Cernis 52  with an equivalent width of 4$\pm$1 m\AA. This  is ascribed to interstellar lithium and appears at a velocity  consistent with that of the much stronger Na I D doublet interstellar absorption lines measured towards the same star. The Na I doublet shows that the interstellar absorption towards Cernis 52 arises mostly  in a discrete individual cloud. There appear to be two competing identifications for the broad 6707 \AA\ feature: a stellar Li\,{\sc i} line or an interstellar  band from the naphthalene cation.

 As shown by the spectrum of HIP 86032 (and other A stars), the 6707 \AA\ region is devoid of stellar
lines, the Li\,{\sc i} is weak or absent because the Li is predominantly
singly-ionized in these atmospheres and its low abundance may be depleted by
mixing. Nonetheless, there are exceptional stars in which the Li\,{\sc i}
doublet is unexpectedly present: is Cernis 52 such an example?. To check this
possibility, we synthesized the spectrum using the code SYNTHE (Kurucz et al.
2005, Sbordone et al. 2005) for a model atmosphere with an effective temperature
of 8500 K, surface gravity of $\log$ g= 4.27, solar metallicity, and a range of Li abundances. The spectrum was broadened for a rotational velocity of 80 km s$^{-1}$. A Li abundance of $\log [N({\rm Li})/N({\rm H})]+12 =4.2$ provides the closest fit to the observed feature, see Fig. 1  with the stellar radial velocity set by the Ca\,{\sc i} 6717 \AA\ and other lines. This abundance is 0.9 dex greater than the standard cosmic  Li abundance of  $\log [N({\rm Li})/N({\rm H})]+12
=3.3$  and such a high Li overabundance is quite exceptional. The sole example among stars of the upper main sequence is apparently a star in the open cluster NGC
6633 (Deliyannis, Steinhauer, \& Jeffries 2002) with a  Li abundance of 4.3 but of a lower effective temperature ($T_{\rm eff} = 7086$ K). The predicted  stellar Li\,{\sc i} profile is not a satisfactory fit to the observed profile and although we cannot definitively exclude at present the possibility for a remarkable Li overabundance in the star  we do not consider this a likely explanation for the 6707 \AA\  feature. Subordinate Li\,{\sc i} lines at 6104 \AA\ and 8126 \AA\ would not be expected to be detectable in our spectra of  Cernis 52.

\begin{figure}
\includegraphics[angle=0,width=8.5cm]{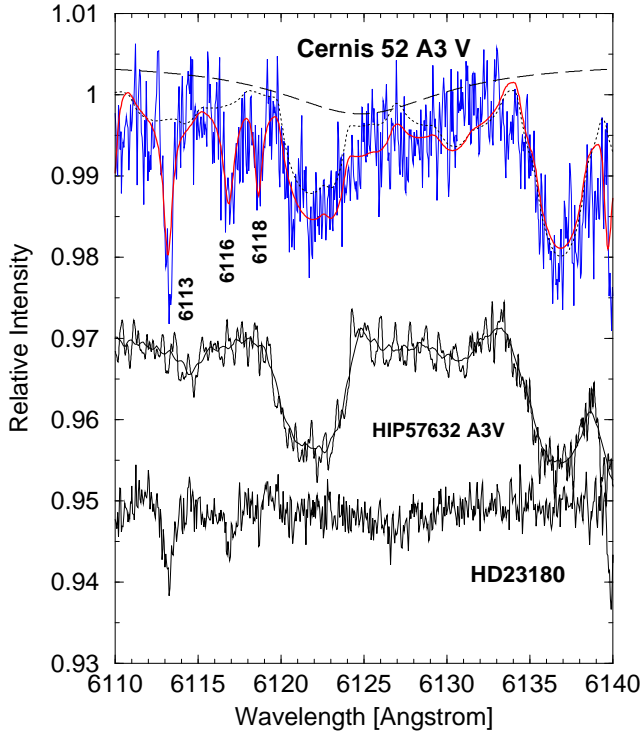}
\caption{Spectrum of star Cernis 52 in the region of the third most intense
 laboratory band of the naphthalene cation (thin blue line). We plot (thick red line) a combination of a synthetic spectrum (same model than in previous figures) with a Lorentzian of width 12 \AA~ at the expected location for this band and simulation of known DIBs with positions at 6613.18, 6616.84 and 6118.63 \AA. A comparison is made with the spectrum of a similar faster rotating star  HIP 57632 (from Allende-Prieto et al. 2004)   and with our spectrum of the reference star HD 23180.}
\end{figure}

 Laboratory spectra of the naphthalene cation have been obtained under conditions relevant to the interstellar medium (Pino, Boudin \& Br\'echignac 1999; Romanini et
 al. 1999; Biennier et al. 2003). These place the strongest band of the  D$_2$$\leftarrow$D$_0$ system  at 6707.0 \AA~ with a precision of about 0.5 \AA~ and with a
 Lorentzian profile with a FWHM of 12$\pm1$ \AA.  In Fig. 1 we include a fit that reproduces the observed spectrum  (solid line at the top spectrum) using a
 Lorentzian profile of width 12 \AA~ centered at 6707.4 \AA\ with superimposed profiles for the known DIBs (wavelengths and widths adopted from Hobbs et al. 2008), interstellar lithium and the photospheric bands. The Lorentzian has an equivalent width of 200 m\AA.

In the search for supporting evidence for this identification of the leading band of the naphthalene cation, we searched for weaker bands.   According to laboratory measurements (Biennier et al. 2003),  the four strongest naphthalene cation absorption peaks are  at 6707.0, 6488.9, 6125.2 and 5933.5 \AA~with a precision of about  0.5
 \AA. Along the  series,  the  intensity decreases as we move to  shorter wavelengths, with a ratio of a factor 2 between each two consecutive  bands  in wavelength. In Fig. 2 we plot the spectrum
 of Cernis 52 in the  region of the second most intense transition in the visible spectrum of  C$_{10}$H$_{8}^+$ in the gas phase. This spectral region is dominated by the presence of a very broad 
 band  at 6494.2 \AA~(Jenniskens et al.  1994), which results from the combination of photospheric absorptions  with
 a rather  broad DIB at 6494.05 \AA~ listed  by Hobbs et al. (2008). The location of the stellar continuum in Fig. 2 is affected by this broad DIB whose strength is
 difficult to estimate in Cernis 52 and the blue wing of stellar H$\alpha$ which makes apparent a progressive decline of the continuum towards the red.  We  detect a new band at 6488.7  \AA~in the
 blue  wing of the broad photospheric feature.  This band appears   partly contaminated by a narrower (0.9\AA) known DIB at 6489.47 \AA~   (Hobbs et al. 2008). At the top of Fig. 2 we plot superimposed on the observed spectrum  a combination of a synthetic photospheric spectrum with a Lorentzian of width 12 \AA~ centered at 6488.7 \AA~ and known DIBs in this spectral region simulated using wavelengths and widths from Hobbs et al. (2008).  From the area of the Lorentzian we infer an upper limit to the equivalent width of the second  naphthalene cation band of 100 m\AA.

 The wavelengths and relative strengths of the 6488 and the 6707 \AA~bands are consistent with those of  the first two transitions of the D$_2$$\leftarrow$D$_0$ system of C$_{10}$H$_{8}^+$.
  In Fig. 3 we can see a weak feature
 coincident in position with the third most intense transition of the naphthalene cation at 6125.2 \AA. Unfortunately, this is
 superimposed to the red wing of a stellar photoshperic feature as shown in the synthetic spectrum  plotted in the same  figure. We combine the synthetic photospheric spectrum with a Lorentzian of
 12 \AA~ (intensity scaled as indicated above) and with a simulation of known DIBs (parameters adopted from Hobbs et al. 2008, as in the previous cases).   The good fit of the combined spectrum  suggests  that indeed the third band of the naphthalene cation could be present  with an equivalent width of $\sim$ 50 m\AA.    Comparison with  similar spectral type stars and the reference star HD 23180 shows no   telluric or photospheric  bands at this wavelength. 
 Searching for weaker transitions of this cation will require much higher quality data.

Adopting  for the oscillator strength of the transition at 6707 \AA~ a value of f=0.05 (Pino et al. 1999) and using the  equivalent width of the Lorentzian fit, we derive a probable C$_{10}$H$_{8}^+$ column density of  N$_{np}^+$= 1 x 10$^{13}$cm$^{-2}$ from the Spitzer formula (with an uncertainty of 60 \%). Assuming a ratio C/H= 3.7 x 10$^{-4}$ and hydrogen column density per unit interstellar absorption N$_H$/A$_V$ = 2 x 10$^{21}$ cm$^{-2}$ (with A$_V$ =3 E(B-V) )  we calculate that 0.008 \% of the total carbon in the intervening cloud could be in the form of naphthalene cations.

In diffuse clouds  and
dense cloud envelopes that are exposed to the hard interstellar radiation PAHs are expected to be  ionized (Wagner, Kim, \& Saykally 2000; Mattioda,  Hudgins, \& Allamandola  2005). This will be the case of naphthalene with a low ionization potential of 8.1 eV.  A recent
computational study of H-addition reactions to the naphthalene cation shows that most of these reactions have little to no barrier and that hydrogen atoms prefer to join the same ring until it is
saturated rather than distribute themselves randomly among the rings (Ricca,  Bakes, Bauschlicher 2007). Hydrogen additions are expected to occur in energetic regions where the molecule is ionized
and there is a  large amount of hydrogen available.  The hydrogen  additions to the naphthalene cation will break the symmetry
of the original naphthalene, leading to an effective dipole moment. In the physical conditions of the clouds these molecules are spinning very fast, with rotational frequencies of tens of GHz, and
will radiate in the microwave range. We have produced a simple model of the electric dipole emission of these particles to estimate its contribution to the anomalous microwave emission detected in
this line of sight. We assume the abundance of naphthalene cations derived above, a dipole moment for the hydrogenated cations comparable to that  of the CH bond (0.3 Debye) and the typical 
physical conditions of a Perseus dense molecular cloud (Iglesias-Groth 2005, 2006). We predict that hydrogenated naphthalene cations would contribute in the range  1-5 x 10$^{-21}$ Jy cm$^{2}$ sr$^{-1}$H$^{-1}$ at the peak frequency of  80 GHz.  This is  just a very small fraction of the total energy emitted in the microwave range by the cloud but illustrates  the role  that thermodynamically stable protonated PAHs might play in the anomalous microwave emission.

\section{Conclusions}

We have recorded the spectrum of star Cernis 52 between 3800 and  10000 \AA~ with resolving power higher than 40000.  In  the range  5800-6800 \AA, we detect   features of likely interstellar origin which are  consistent in wavelength and strength ratios with the  strongest  laboratory  gas-phase transitions of the naphthalene cation. These   features and other DIBs are detected in the line of sight toward  a region of the Perseus molecular complex  with previously reported  anomalous microwave emission.  To the best of our knowledge the new  bands have not been previously detected as DIBs.This suggests that the carrier is either not ubiquituous or much less abundant  in the  general interstellar medium than in the line of sight under  investigation.  The tentative identification of the cation of the simplest PAH reinforces the hypothesis  that this class of molecules could be  carriers of the anomalous microwave emission and encourages accurate laboratory gas-phase  measurements of other  PAHs. 

In  molecular clouds naphthalene is expected to condense onto refractory dust grains.  A large variety of ice mantles can form, but H$_{2}$O will very probably be a dominant species in the ices (Sandford 1996). Recent studies have shown how, under astrophysical conditions,  UV radiation on naphthalene in H$_{2}$O ice produces naphthoquinones. These molecules, if further functionalized with a methyl  group and a long isoprene chain, play essential roles in biochemical functions (Bernstein et al. 2001). Since quinones and aromatic alcohols  are already found in meteoritic material, the detection of naphthalene in molecular clouds shows a possible path for these  biogenic compounds from the chemistry of clouds in which new stars form to the pre-biotic molecules present in the  protoplanetary material.

\acknowledgments
We thank R. Watson and R. Rebolo for many valuable discussions and B.E. Reddy for help with  spectroscopic observations.  DLL thanks the Robert A. Welch Foundation of Houston for support.J.I.G.H. acknowledges support fromthe eU copntract MEXT-CT-2004-014265 
(CIFIST).  This work has been partially founded   by project AYA2007-64748 of the Spanish Ministry of Education and Science.


\begin{thebibliography}{}
\bibitem[]{ Allama85} Allamandola, L. J., Tielens,   A. G. G. M., \& Barker, J. R. 1985 , \apjl~ 290, L25
\bibitem[]{Allama89} Allamandola, L. J., Tielens,G. G. M., \& Barker, J.R. 1989,  ApJS 71, 733
\bibitem[]{Allende04} Allende Prieto, C., Barklem, P.S., Lambert, D.L., Cunha, K. 2004, A\&A 420, 183
\bibitem[]{Bernstein01} Bernstein, M.P, Dworkin,  J.P., Sandford, S.A. \& Allamandola, L.J. 2001,  Meteor. Plane. Sci. 36, 351
\bibitem[]{Biennier03} Biennier, L., Salama, F., Allamandola, L., Scherer, J. 2003, J. Chem. Phys. 118, 17, 7863
\bibitem[]{cernis93} Cernis, K. 1993, Baltic Astronomy 2, 214
\bibitem[]{deliyannis} Deliyannis, C.P., Steinhauer, A. \& Jeffries, R.D. 2002, \apjl, 577, L39
\bibitem[]{drain98a} Draine, B.T. \& Lazarian, A. 1998a, \apjl, 494, L19
\bibitem[]{galaz01} Galazutdinov, G.A., Musaev, F.A., Kre{\l}owski, J. \& Walker, G.A.H. 2000, PASP, 112, 648
\bibitem[]{gratton01} Gratton, R. G., Bonanno, G., Bruno, P., Cali, A., Claudi, R. U., Cosentino, R.,
Desidera, S., 2001, Experimental Astron., 12,
\bibitem[]{herb95} Herbig, G.H., 1995, Ann. Rev. Astron. Astrophys. 33, 19 
\bibitem[]{Hilde07} S. R. Hildebrandt,S. R, Rebolo, R., Rubiño-Mart\'\i n, J.A., Watson, R.A., Guti\'errez, C.M., Hoyland, R.J., Battistelli,
E.S. 2007, MNRAS 382, 594
\bibitem[]{hobbs08} Hobbs, L. M., York, D. G., Snow, T. P., Oka, T., Thorburn, J.
A.; Bishof, M., Friedman, S. D., McCall, B. J. et al. 2008, arXiv:astro-ph/0804.2094H
\bibitem[]{ig05} Iglesias-Groth, S. 2005, \apjl~ 632, L25
\bibitem[]{ig06} Iglesias-Groth, S. 2006, MNRAS 368,1925
\bibitem[]{jen94} Jenniskens, P. \& D\'esert, F.X. 1994, A\&AS 106, 39
\bibitem[]{kre01} Kre{\l}owski, J., Galazutdinov, G.A., Musaev, F.A. \& Nirski, J. 2001, MNRAS 328, 810
\bibitem[]{kre95} Kre{\l}owski, J. \& Sneden, C. 1995 in The Diffuse Interstellar Bands, ed. A.G.G. M. Tielens and T. P. Snow (Dordrecht, Kluwer) 13.
\bibitem[]{kre86} Kre{\l}owski, J., Walker, G. A. H. 1987, \apj~ 87, 312
\bibitem[]{kre01} Kre{\l}owski, J., Galazutdinov, G. A., Musaev, F. A., Nirski, J.J. 2001, MNRAS 328, 810
\bibitem[]{kogut96} Kogut, A., Banday, A. J., Bennett, C. L., Gorski, K. M., Hinshaw, G.\& Reach, W. T. 1996, \apj~ 460, 1]
\bibitem[Kurucz(2005)]{2005MSAIS...8...14K} Kurucz, R.~L.\ 2005, Memorie della Societa Astronomica Italiana Supplement, 8, 14	
\bibitem[]{leger85} L\'eger, A., \& L.  d\'~Hendecourt, L. 1985,  A\&A  146, 81
\bibitem[]{leitch97} Leitch, E. M., Readhead, A. C. S., Pearson, T. J. \& Myers, S. T. 1997, \apjl~486, L23
\bibitem[]{Mattioda05} Mattioda, A. L., Hudgins, D. M. $\&$ Allamandola, L. J. 2005, \apj~ 629, 1188
\bibitem[]{Mulas06} Mulas, G. Malloci, G., Joblin, J.  $\&$  Toublanc, D. 2006, A\&A  460, 93
\bibitem[]{oliv99} de Oliveira-Costa, A.,  Tegmark, M., Guti\'errez, C. M.,  Jones, A.  W.,  Davies, R. D., Lasenby, A. N.,  Rebolo, R., \& Watson, R. A.  1999,  ApJ 527, L9
\bibitem[]{oliv02} de Oliveira-Costa, A. et al. 2002, \apj~ 567, 363
\bibitem[]{pino99} Pino, T., Boudin, N. $\&$ Br\'echignac, P. 1999, J. Chem. Phys 111, 16, 7337
\bibitem[]{puget89} Puget, J. L. $\&$ L\'eger, A. 1989,  ARA\&A  27, 161
\bibitem[]{ricca07} Ricca, A., Bakes, E. L. O., Bauschlicher, C. W. 2007, \apj~  659, 858
\bibitem[]{romanini99} Romanini, D., Biennier, L., Salama, F., Kachanov, A., Allamandola, L.J. $\&$ Stoeckel, F. 1999, Chem. Phys. Lett. 303, 165. 1
\bibitem[]{Sandford96} Sandford, S. 1996, ASPC 97, 29
\bibitem[]{Sarre06} Sarre, P.J. 2006, J. Mol. Spectr. 238, 1
\bibitem[Sbordone(2005)]{2005MSAIS...8...61S} Sbordone, L. 2005, Mem. Soc. Astron. It. Suppl., 8, 61
\bibitem[]{Snow92} Snow, T.P. 1992, \apj~ 401, 775
\bibitem[]{taylor03} Taylor, A. C. et al. 2003, MNRAS 341, 1066
\bibitem[]{tull95} Tull, R. G., MacQueen, P. J., Sneden, C., Lambert, D. L. 1995, PASP, 107, 251
\bibitem[]{VanderZwt85} Van der Zwet,G. P., Allamandola, L. J. 1985,  A\&A 146, 76
\bibitem[]{wagner00} Wagner, D. R., Kim, HS., Saykally, R. J.  2000, \apj~  545, 854
\bibitem[]{watson91} Watson,R.A. et. al 2005,  \apj~ 624, L89
\bibitem[]{watson91} Watson, R. A. et al., in "CMB and Physics
of the Early Universe" 20-22 April 2006, Ischia, Italy, p.66


\end{thebibliography}
\end{document}